\documentclass{ismdproc}

\begin{document}
\title{Proton Structure from HERA to LHC}


%

%

%
\author{{\slshape Amanda Cooper-Sarkar}  for the H1 ans ZEUS Collaborations\\[1ex]
University of Oxford, Denys Wilkinson Bdg, Keble Rd, Oxford OX1 3RH, GB }

\contribID{xy}  
\confID{yz}
\acronym{ISMD2010}
\doi            

\maketitle

\begin{abstract}
  The kinematic region covered by the LHC experiments probes low
values of Bjorken-$x$. At the scale relevant for $W$ and
 $Z$ production the central rapidity region covers 
$10^{-3} < x < 10^{-1}$, 
for $\surd{s}=7~$TeV. This means that it is the HERA data which give the most 
insight into the behaviour of parton distribution functions (PDFs) 
for the early phase of running at the LHC. 
The H1 and ZEUS experiments are combining their data so as to provide a legacy 
of HERA results. The present status of the data combinations and the impact of 
these data on our knowledge of parton distribution functions is explored.
\end{abstract}

\section{Introduction}
 H1 and ZEUS have combined various sub-sets of their data.
The combination of inclusive cross section data from HERA-I and the 
PDF fit based on these data are already 
published~\cite{h1zeus:2009wt}. In 2010 further data have been combined and 
PDF fits to the augmented data sets have been made available in 
preliminary form. In Sec.~\ref{sec:inc1} results from the published 
combination are reviewed. In Sec.~\ref{sec:charm} results from a combination of
$F_2^{c\bar{c}}$ data are presented and their impact on predictions for 
$W$ and $Z$ production at the LHC discussed. 
In Sec.~\ref{sec:lowenergy} results from 
the combination of inclusive cross section data taken at lower proton beam 
energies are discussed. Finally, in Sec.~\ref{sec:inc2} an updated combination
of all inclusive data from HERA-I and HERA-II running is shown and 
predictions for the LHC $W$ and lepton asymmetries, from a PDf fit to these 
data, are presented.
\section{Results}
\subsection{Inclusive data from HERA-I running}
\label{sec:inc1}
\begin{figure}[tbp]
\vspace{-2.0cm}
\begin{center}
\begin{tabular}{ll}
\psfig{figure=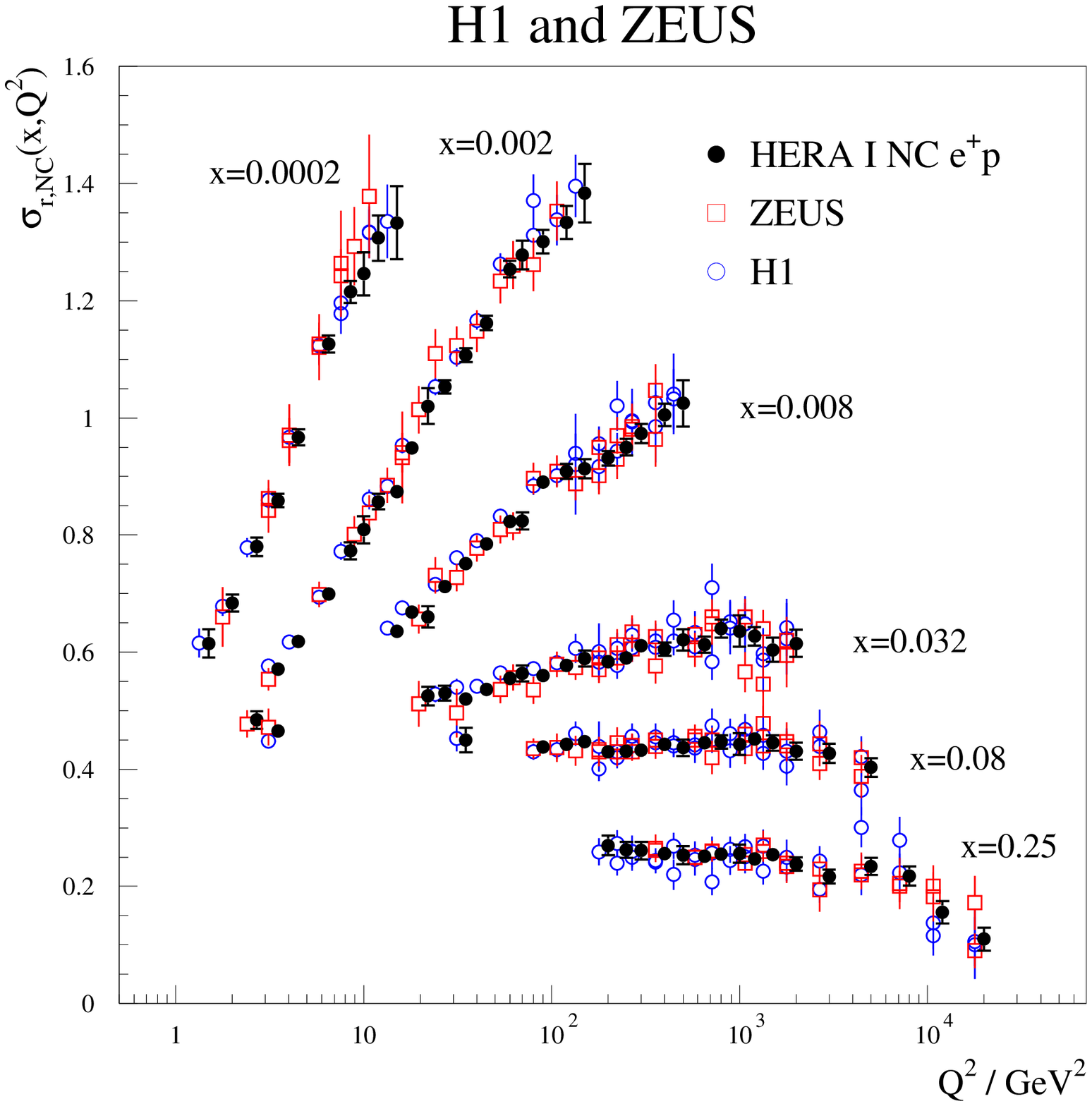,width=0.40\textwidth}~~
\psfig{figure=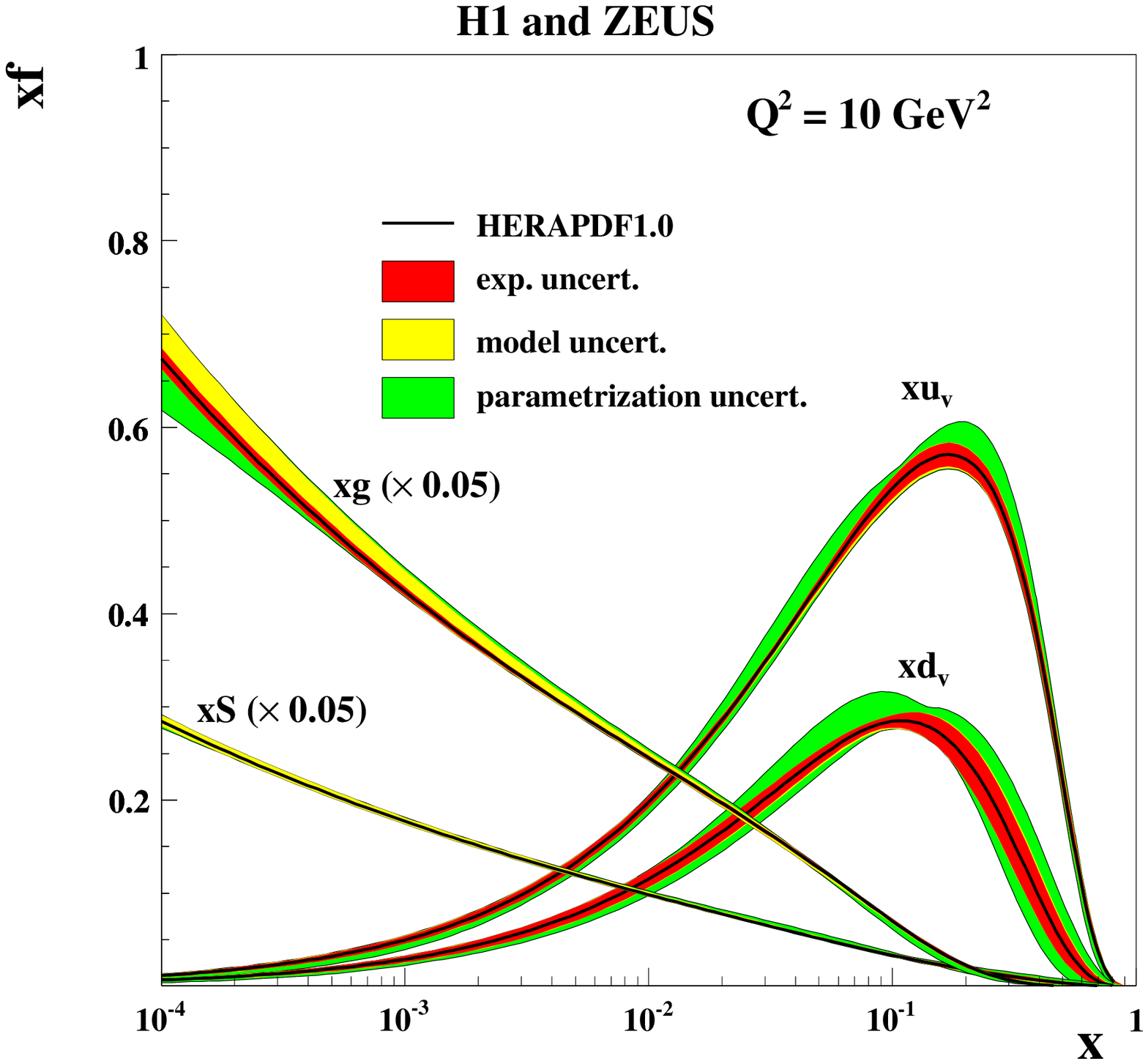,width=0.40\textwidth} \\
\end{tabular} 
\caption{
Left: HERA combined data points for the NC $e^+p$ cross-section as a function of
$Q^2$ in selected bins of $x$, compared to the separate ZEUS and H1 data sets input to the combination.  Right: Parton distribution functions from HERAPDF1.0; $xu_v$, $xd_v$, $xS=2x(\bar{U}+\bar{D})$ and $xg$ at $Q^2=10~$GeV$^2$.}
\label{fig:herapdf10}
\end{center}
\end{figure}

The inclusive cross section data, 
from the HERA-I running period 1992-2001, for Neutral Current (NC) and 
Charged Current (NC), $e^+p$ and $e^-p$ scattering have been 
combined~\cite{h1zeus:2009wt}. The combination procedure pays particular 
attention to the correlated systematic uncertainties of the data sets 
such that resulting combined data benefits from the best features of each 
detector. The combined data set has systematic 
uncertainties which are smaller than its statistical errors and the total 
uncertainties are small ($1-2\%$)
over a large part of the kinematic plane. The combined 
data is compared to the separate input data sets of ZEUS and H1 in 
Fig.~\ref{fig:herapdf10}.

These data are used as the sole input to a PDF fit called the 
HERAPDF1.0~\cite{h1zeus:2009wt}. The motivations for performing a HERA-only 
fit are firstly, that the combination of the HERA data yields a very accurate 
and consistent data set such that the experimental uncertainties on the PDFs 
may be estimated from the conventional $\chi^2$ criterion $\Delta\chi^2=1$.
Global fits which include dats sets from many different experiments often use 
inflated $\chi^2$ tolerances in order to account for marginal consistency of 
the input data sets. Secondly, the HERA data are proton target data so that 
there is no uncertainty from heavy target corrections or deuterium corrections 
and there is no need to assume that $d$ in the proton is the same as $u$ in 
the neutron since the $d$-quark PDF may be extracted from $e^+p$ CC data.
Thirdly, the HERA inclusive data give information  
on the gluon, the Sea and the $u$- and $d$-valence PDFs over a wide kinematic region: 
the low-$Q^2$ NC $e^+p$ cross-section data
are closely related 
to the low-$x$ Sea PDF and the low-$x$ gluon PDF is derived from its scaling 
violations; the high-$x$ $u-$ and $d$-valence PDFs are closely 
related to the high-$Q^2$ NC $e^{\pm}p$,CC $e^-p$ and CC $e^+p$ cross sections, 
respectively; the difference between the high-$Q^2$ $e^-p$ and $e^+p$ 
cross-sections gives the valence shapes down to low $x$, $(x\sim 10^{-2})$. The 
HERAPDF1.0 parton distributions are shown in Fig.~\ref{fig:herapdf10}.
HERAPDF provides model and parametrisation uncertainties on the PDFs as well as
experimental uncertainties. 
Parametrisation uncertainties come from varying the central form chosen for 
the parametrisation to allow for more flexible forms, and from varying the starting 
scale $Q^2_0$ at which the parametrisation is input and $Q^2$ evolution begins.
The model uncertainties come 
from varying: the values of the 
charm and beauty quark masses; the value of the parameter which controls the 
relative size of the strange compared to the light quarks; 
and the minimum value
of $Q^2$ for data entering the fit. 

When the HERAPDF is used to predict $W$ and $Z$ cross sections at the LHC it 
is found that the predictions at central rapidity have small total 
uncertainties $\sim 4\%$. However it is also clear that a major contribution 
to this uncertainty comes from the model uncertainty on the charm mass value.
This can be improved using information from data on $F_2^{c\bar{c}}$.

\subsection{Charm data from HERA-I and II running}
\label{sec:charm}
\begin{figure}[tbp]
\vspace{-2.0cm}
\begin{center}
\begin{tabular}{ll}
\psfig{figure=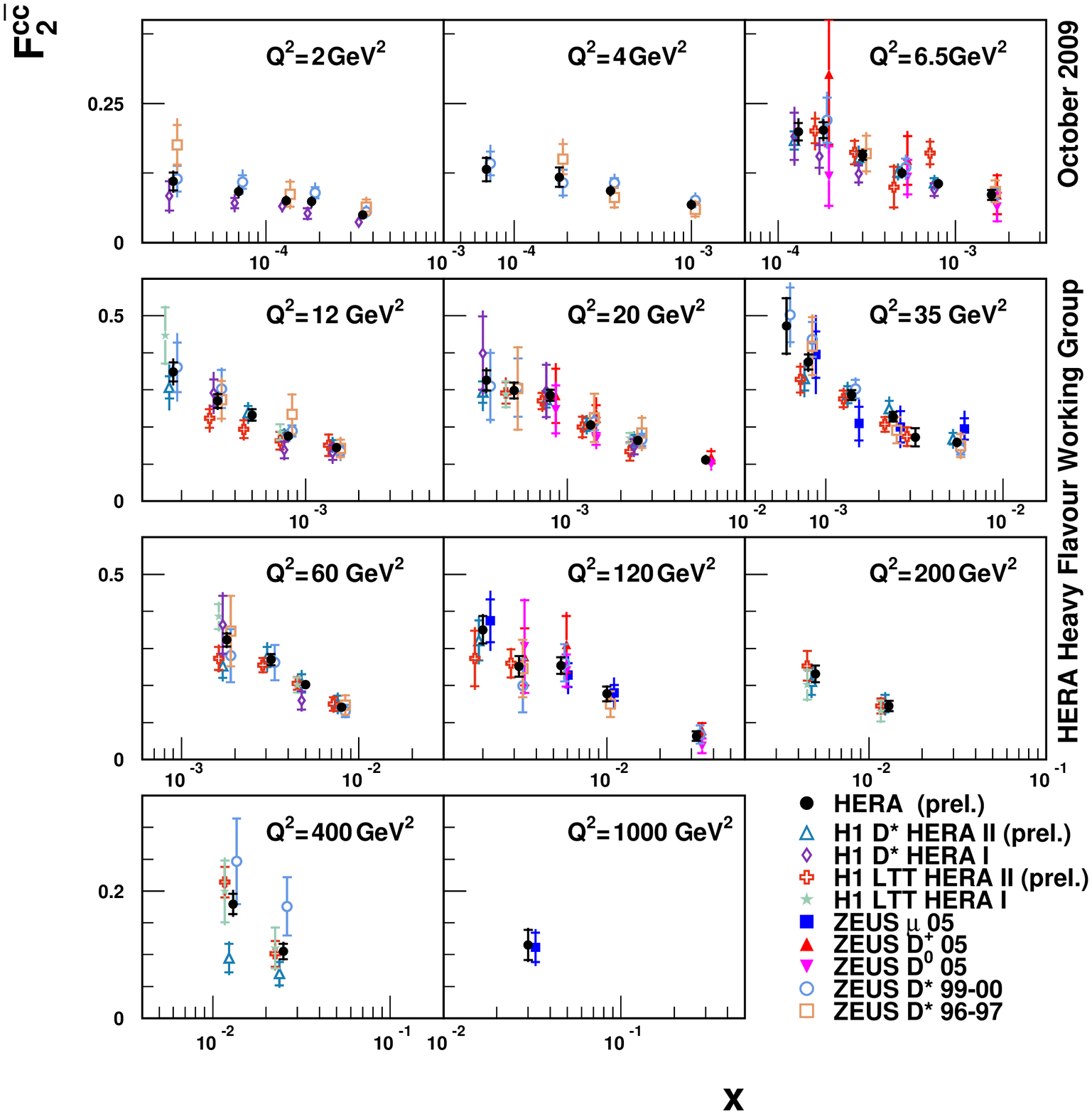,width=0.40\textwidth}~~~~ &
\psfig{figure=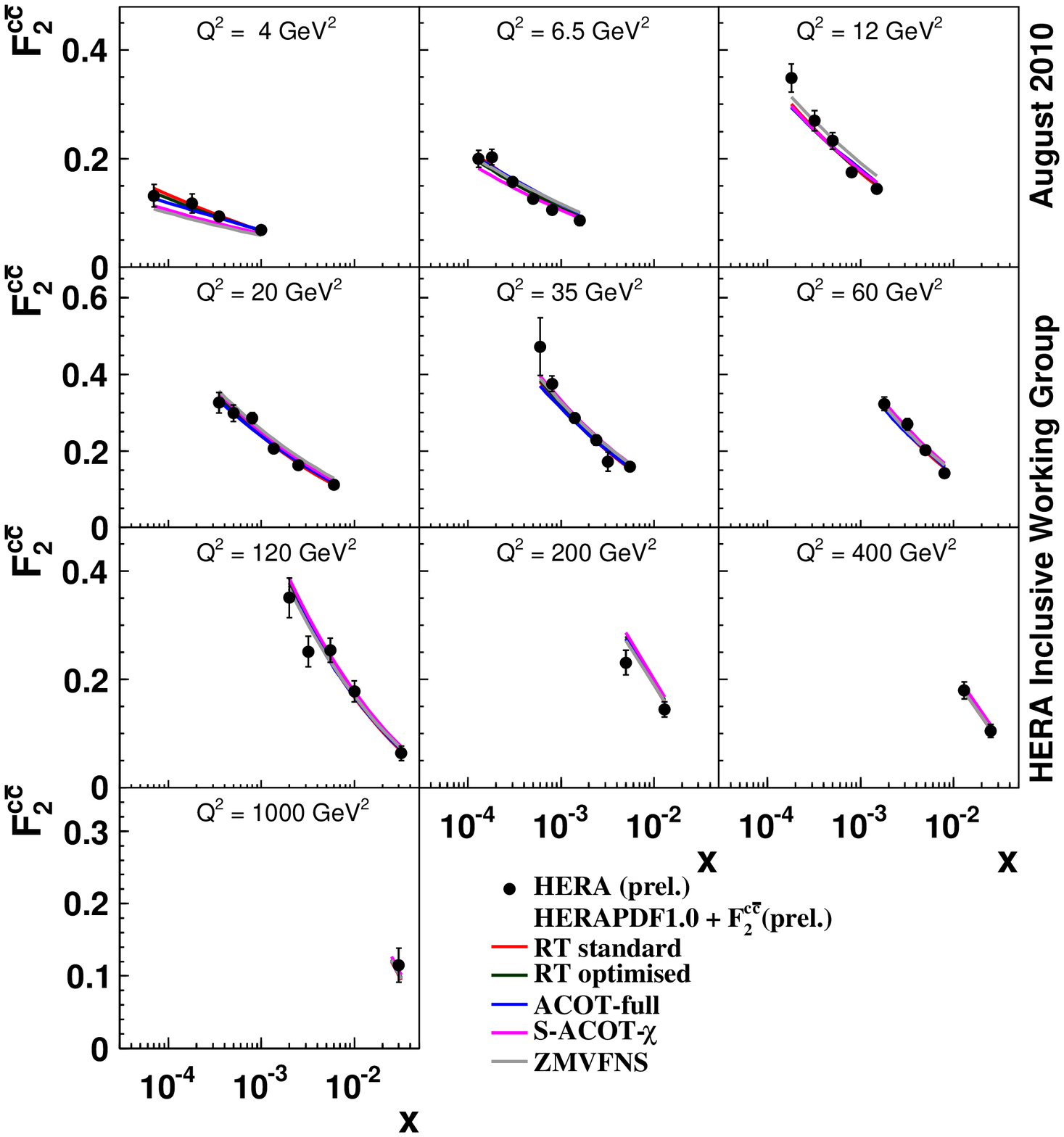,width=0.35\textwidth} \\
\end{tabular} 
\caption{Left: HERA combined data points for $F_2^{c\bar{c}}$, as a function of
$x$ in bins of $Q^2$, compared to the separate ZEUS and H1 data sets input to the combination. 
Right: HERA combined data points for $F_2^{c\bar{c}}$ compared to predictions from the HERAPDF fit to these data plus the HERA-I inclusive DIS data, for various different heavy-quark-mass schemes, as specified in the text.    }
\label{fig:charmdata}
\end{center}
\end{figure}

\begin{figure}[tbp]
\vspace{-2.0cm} 
\centerline{\psfig{figure=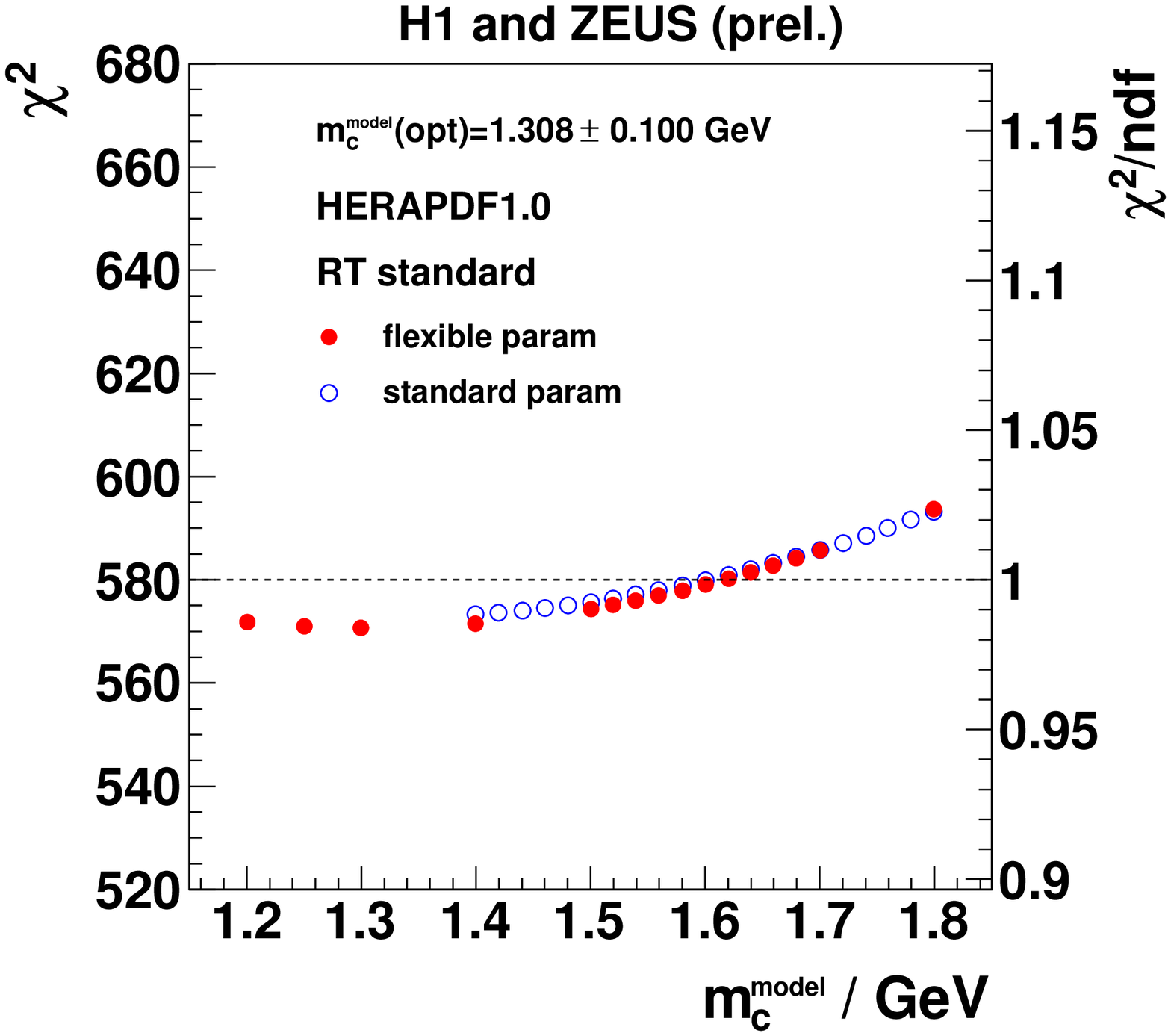,width=0.30\textwidth}~~ 
\psfig{figure=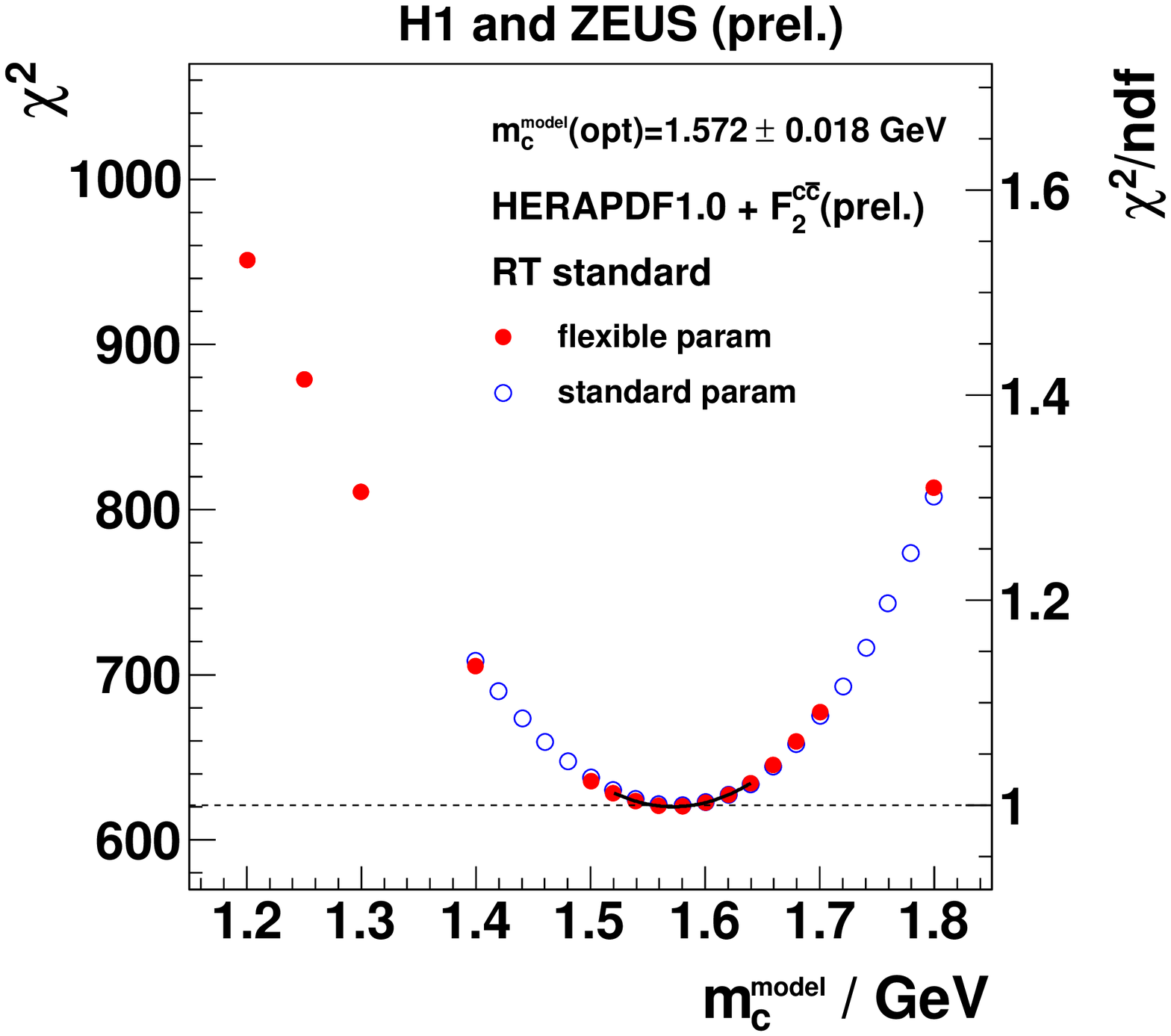,width=0.30\textwidth}}
\centerline{\psfig{figure=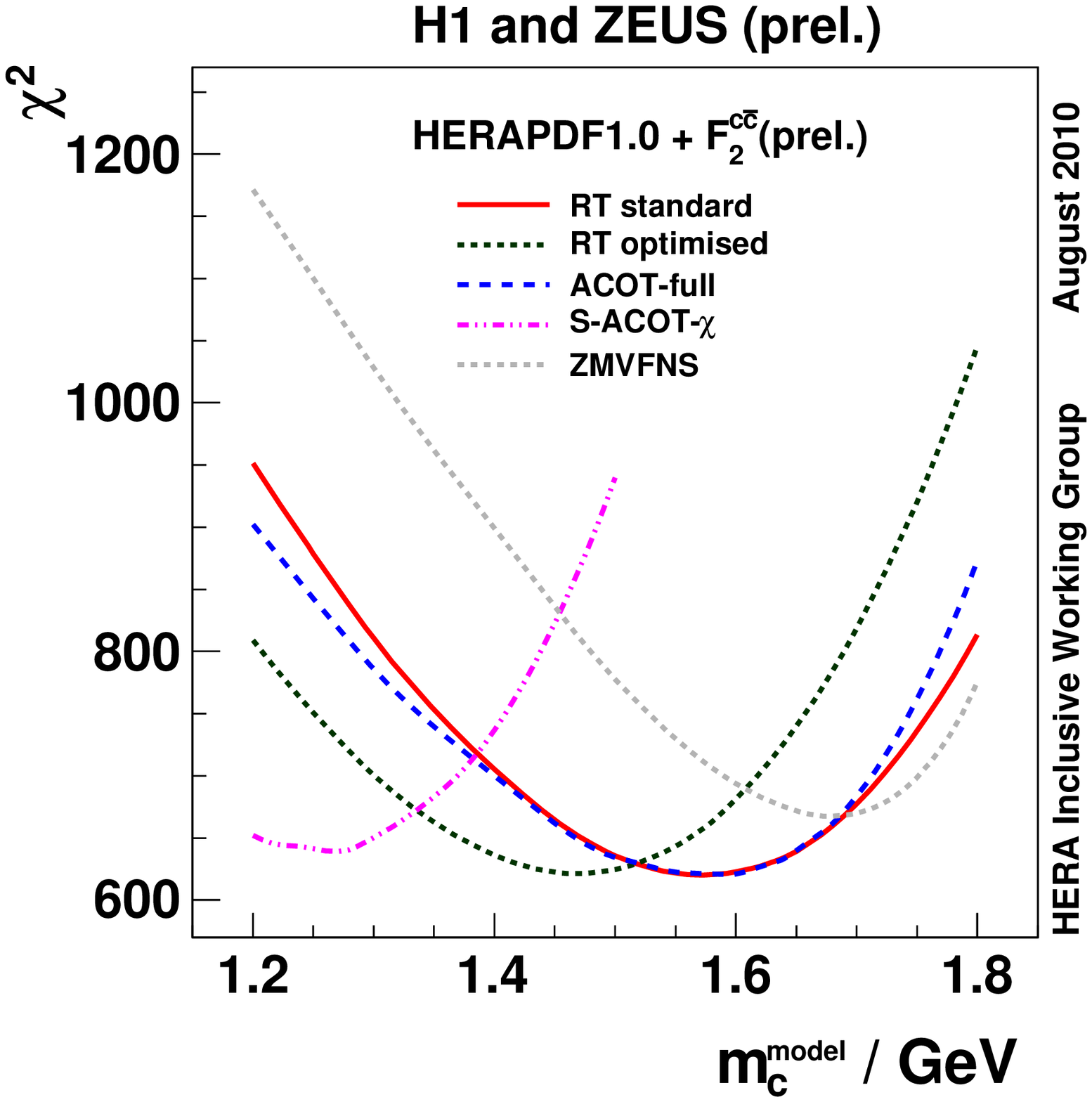,width=0.30\textwidth}~~ 
\psfig{figure=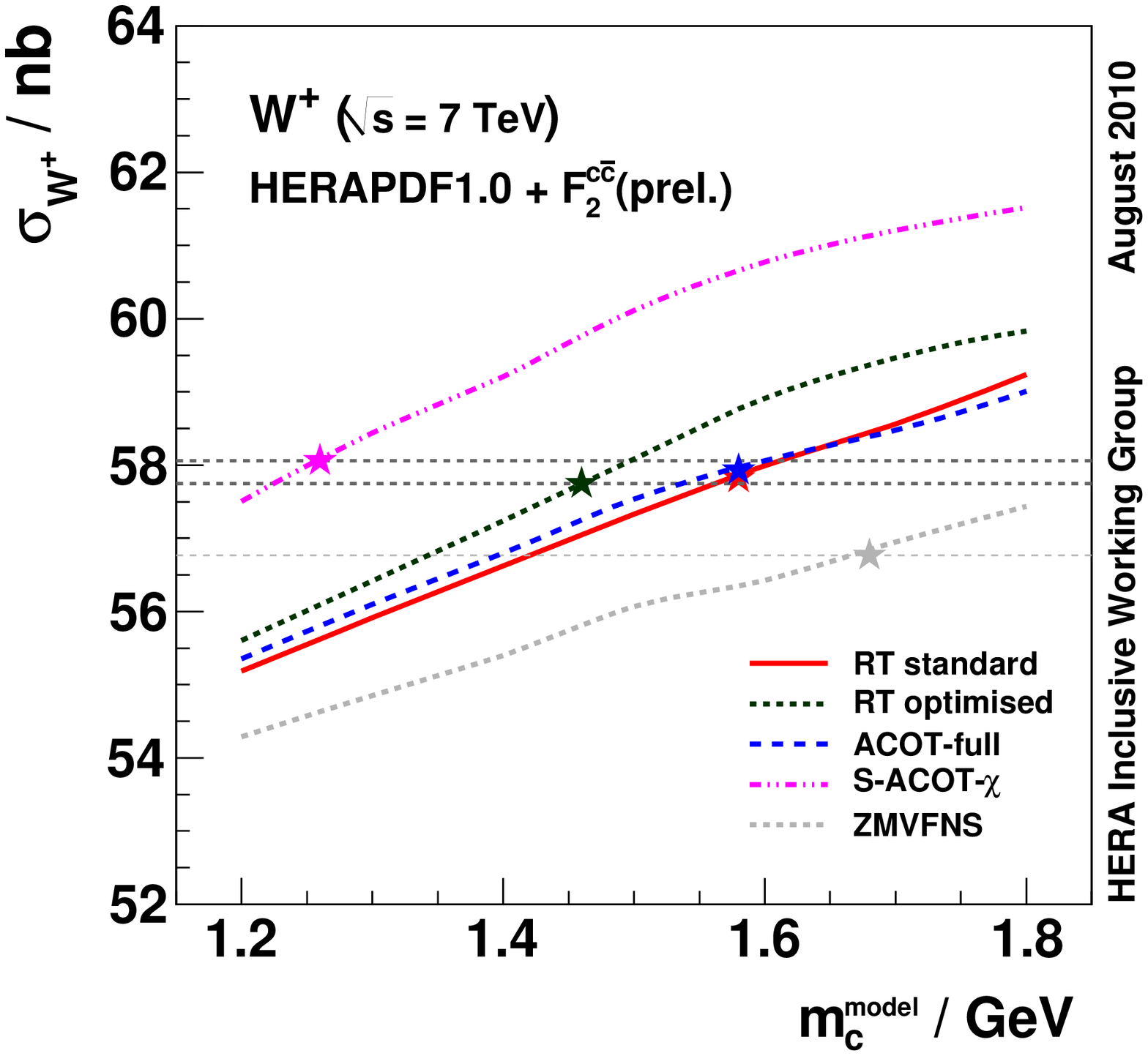,width=0.30\textwidth}}
\caption {The $\chi^2$ of the HERAPDF fit as a 
function of the charm mass parameter $m_c^{model}$. Top left; using the 
RT-standard heavy-quark-mass scheme, when only inclusive DIS data are included 
in the fit. Top right; 
using the RT-standard heavy-quark-mass scheme, 
when the data for $F_2^{c\bar{c}}$ are also included in the fit. Bottom left;
 using various heavy-quark-mass schemes, when the data for $F_2^{c\bar{c}}$
 are also included in the fit. Bottom right: predictions for the $W^+$ cross-sections at the LHC, as a 
function of the charm mass parameter $m_c^{model}$, for various heavy-quark-mass schemes.  
}
\label{fig:charmpred}
\end{figure}

H1 and ZEUS have also combined their data on 
$F_2^{c\bar{c}}$~\cite{charmcomb}.
In Fig.~\ref{fig:charmdata} the combined data are compared to the separate 
data sets which go into the combination.
These data are input to the HERAPDF fit together with the inclusive data 
which were used for HERAPDF1.0. The $\chi^2$ of this fit is sensitive to the 
value of the charm quark mass. Fig,~\ref{fig:charmpred} compares the $\chi^2$, 
as a function of this mass, for a fit which includes these data (top left)
 to that for the
HERAPDF1.0 fit (top right). 
However, it would be premature to conclude that the data 
can be used to determine the charm pole-mass. The HERAPDF formalism uses 
the Thorne-Roberts (RT) variable-flavour-number (VFN) scheme for heavy 
quarks. This scheme is not unique,
specific choices are made for threshold behaviour. In Fig.~\ref{fig:charmpred} 
(bottom left) the $\chi^2$ profiles for the standard and the 
optimized versions (optimized for smooth 
threshold behaviour) of this scheme are compared. The same figure also
 compares the alternative ACOT VFN 
schemes and the Zero-Mass VFN scheme. Each of these schemes 
favours a different value for the charm quark mass, and the fit to the data 
is equally good for all the heavy quark mass schemes (see Fig.~\ref{fig:charmdata} right). However,  the Zero-Mass scheme is $\chi^2$ disfavoured. 
Each of these schemes can be used to predict $W$ 
and $Z$ production for the LHC and their predictions for $W^+$ are shown in 
Fig.~\ref{fig:charmpred} as a function of the charm quark mass. If a particular
value of the charm mass is chosen then the spread of predictions is as large as
$\sim7\%$.
However this spread is considerably reduced $\sim 1\%$ if each heavy quark 
scheme is used at its own favoured
value of the charm quark-mass. Even the Zero-Mass scheme lies only 
$\sim2\%$ below the heavy quark schemes. Furher details of this study are 
given in ref.~\cite{charmstudy}.

\subsection{Low energy proton beam data from 2007}
\label{sec:lowenergy}
\begin{figure}[tbp] 
\begin{center}
\begin{tabular}{ll}
\psfig{figure=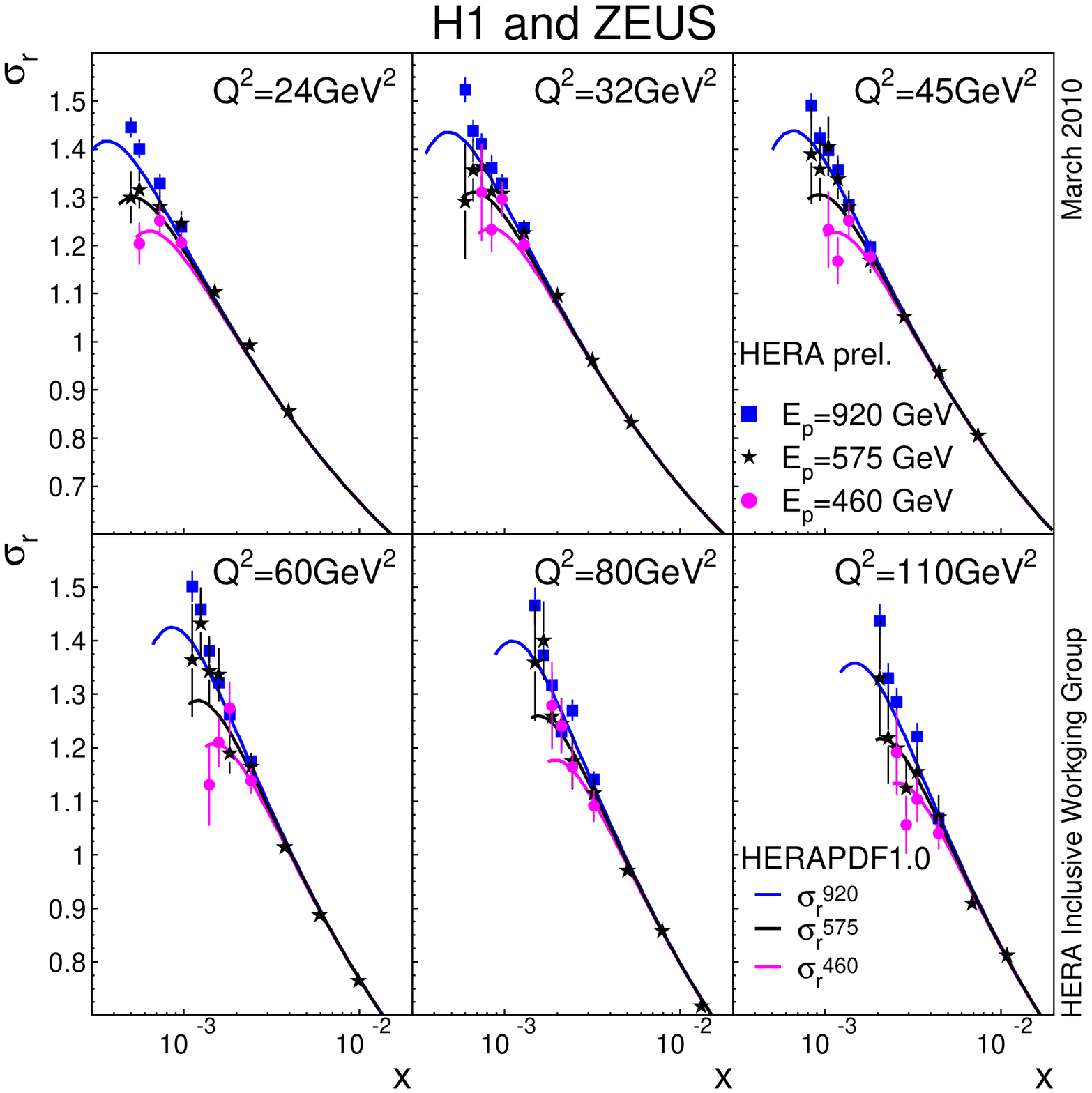,width=0.40\textwidth}~~ &
\psfig{figure=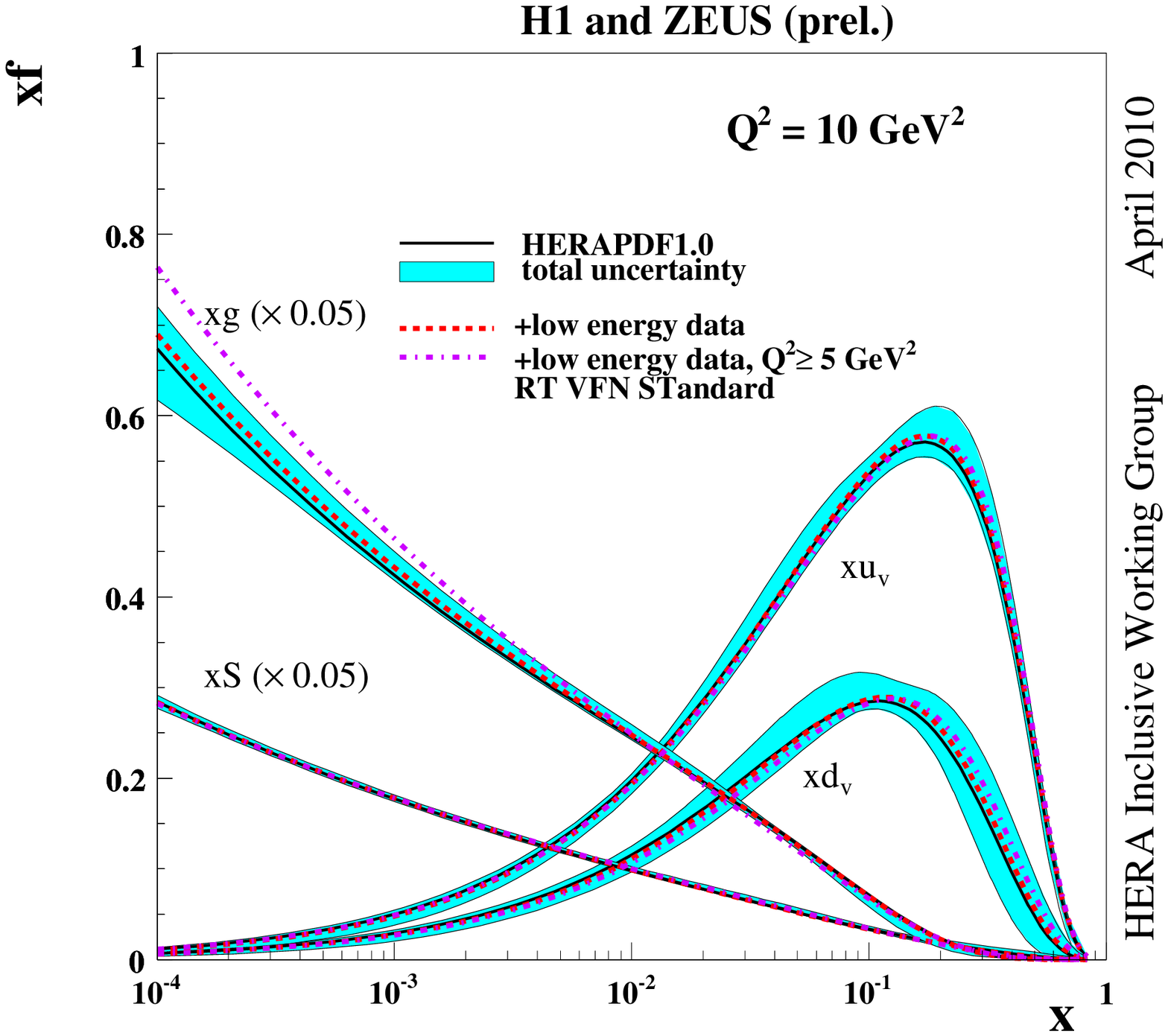,width=0.40\textwidth} \\
\end{tabular} 
\caption{Left: HERA combined data points for the NC $e^+p$ cross-section as a 
function of $x$ in bins of $Q^2$, for three different proton beam energies. 
Right: Parton distribution functions, $xu_v$, $xd_v$, $xS=2x(\bar{U}+\bar{D})$ 
and $xg$ at $Q^2=10~$GeV$^2$, for HERAPDF1.0 and for a HERAPDF fit which also 
includes the low-energy proton beam data, with the standard $Q^2$ cut, 
$Q^2 > 3.5~$GeV$^2$, and for $Q^2 > 5.0~$GeV$^2$.  }
\label{fig:lowenergy}
\end{center}
\end{figure}

In 2007 NC $e^+p$ data were taken at two lower values of the proton beam energy in order
to determine the longiudinal srtucure fuction $F_L$. Some of the H1 and ZEUS 
data sets from these runs have now been combined~\cite{lowenergy} 
and the results for the NC  $e^+p$ cross section are shown in Fig.~\ref{fig:lowenergy}.
These data have been input to the HERAPDF fit together with the inclusive data 
from HERA-I. The resulting parton distributions are compared with those of 
HERAPDF1.0 in Fig.~\ref{fig:lowenergy}. The low energy data are sensitive 
to the choice of minimum $Q^2$ (standard cut $Q^2 > 3.5~$GeV$^2$) 
for data entering the fit. If a 
somewhat harder cut, $Q^2 > 5~$GeV$^2$, is made, a steeper gluon distribution 
results - see Fig.~\ref{fig:lowenergy}, whereas for the HERAPDF1.0 this 
variation 
of cuts results in PDFs which lie within the PDF uncertainty bands. This 
sensitivity is also present 
if an $x$ cut, $x > 5\times 10^{-4}$, or a 'saturation inspired' cut,
$Q^2 > 0.5~x^{-0.3}$, is made. This sensitivity may indicate the breakdown of 
the DGLAP formalism at low $x$~\cite{lowestudy}.

\subsection{High-$Q^2$ data from HERA-II running}
\label{sec:inc2}
\begin{figure}[tbp]
\vspace{-2.0cm} 
\begin{center}
\begin{tabular}{ll}
\psfig{figure=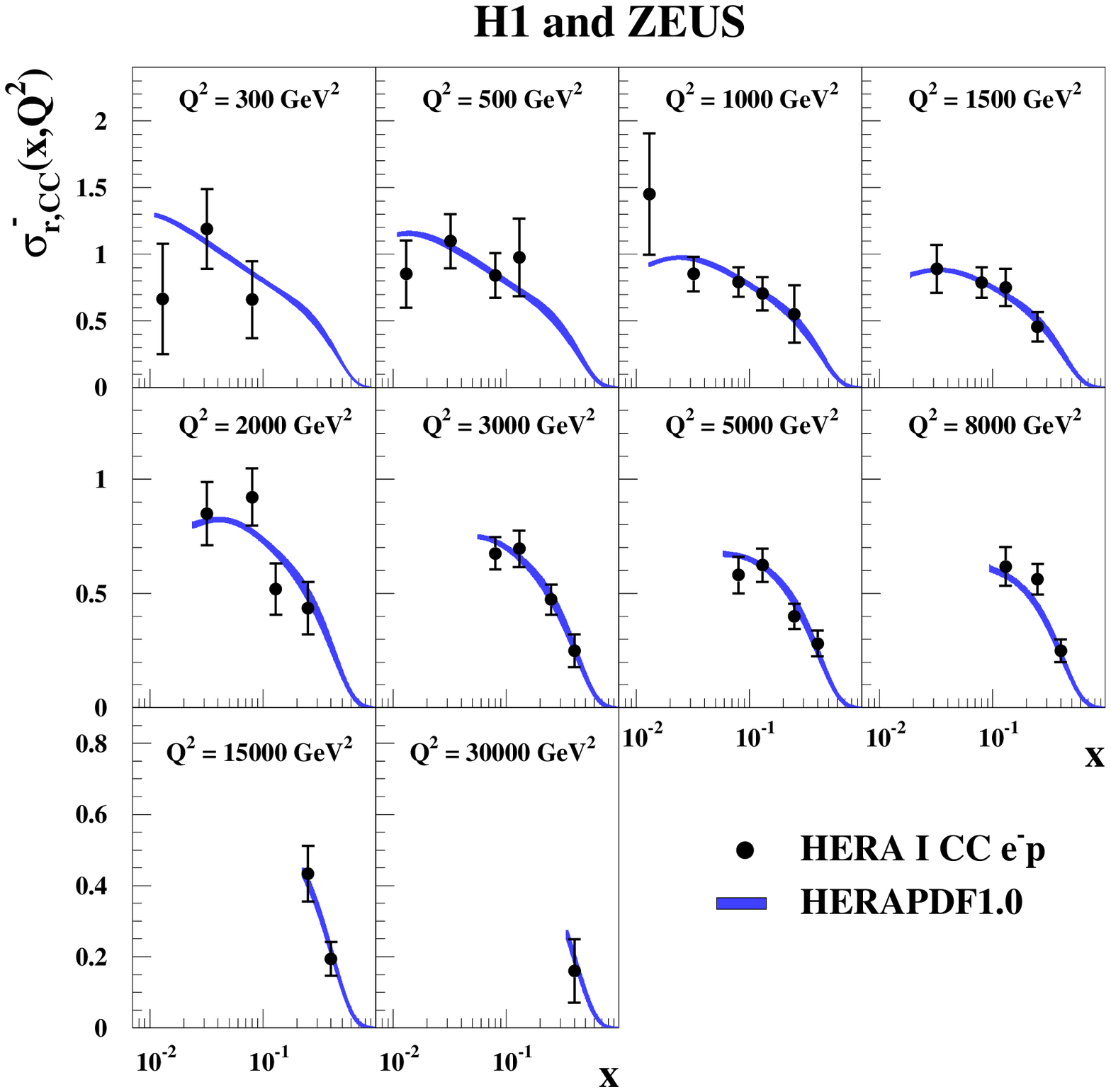,width=0.40\textwidth}~~ &
\psfig{figure=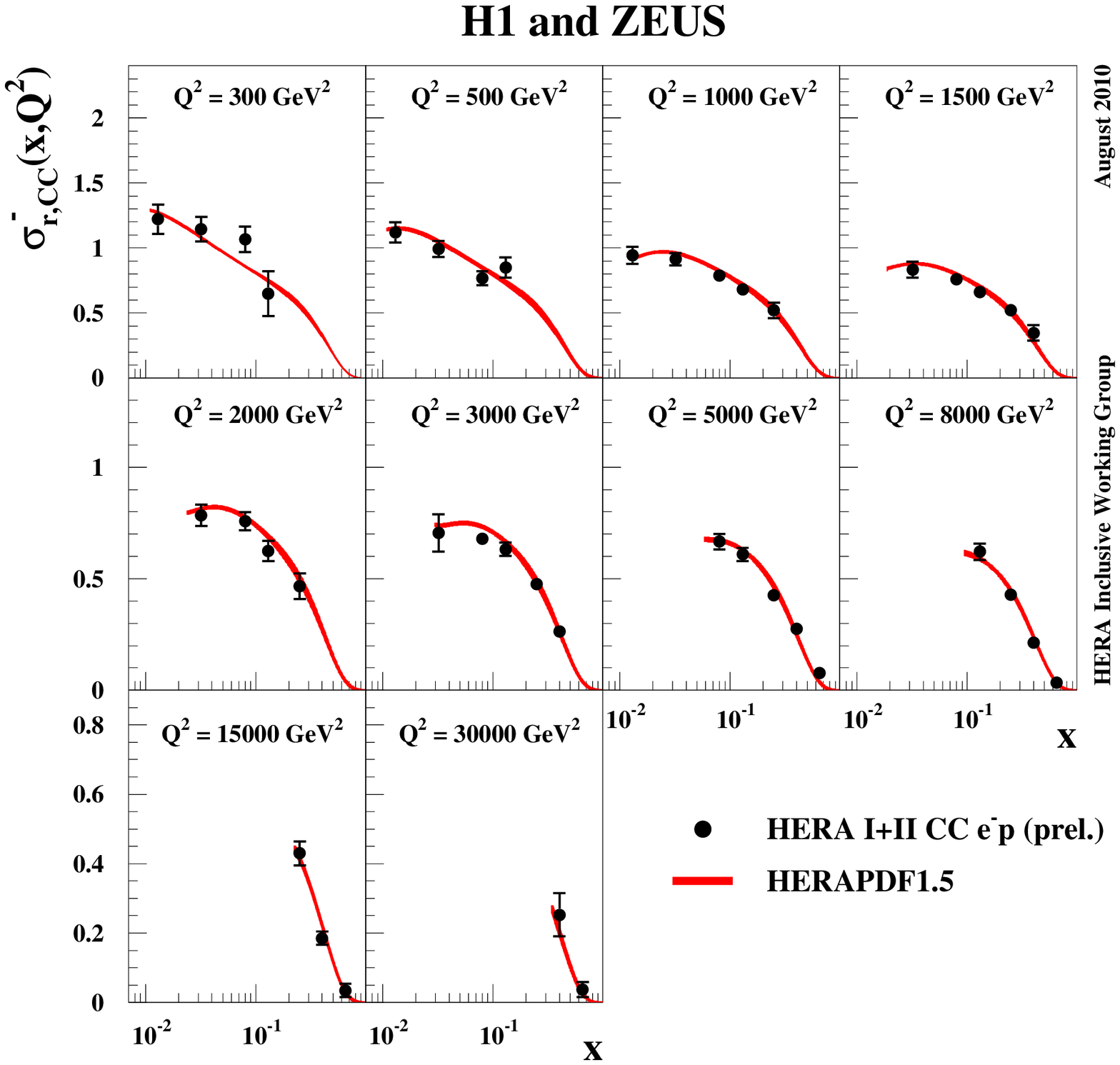,width=0.40\textwidth} \\
\end{tabular} 
\caption{HERA combined data points for the CC $e^-p$ cross-section as a  
function of $x$ in bins of $Q^2$. Left: from the HERA-I run period. Right: 
from the HERA-I and II run periods. On each plot the HERAPDF fit which includes
the corresponding data is illustrated: the HERAPDF1.0 fit on the left hand plot and the
HERAPDF1.5 on the right hand plot.}
\label{fig:ccem1015}
\end{center}
\end{figure}

\begin{figure}[tbp]
\vspace{-2.0cm} 
\begin{center}
\begin{tabular}{ll}
\psfig{figure=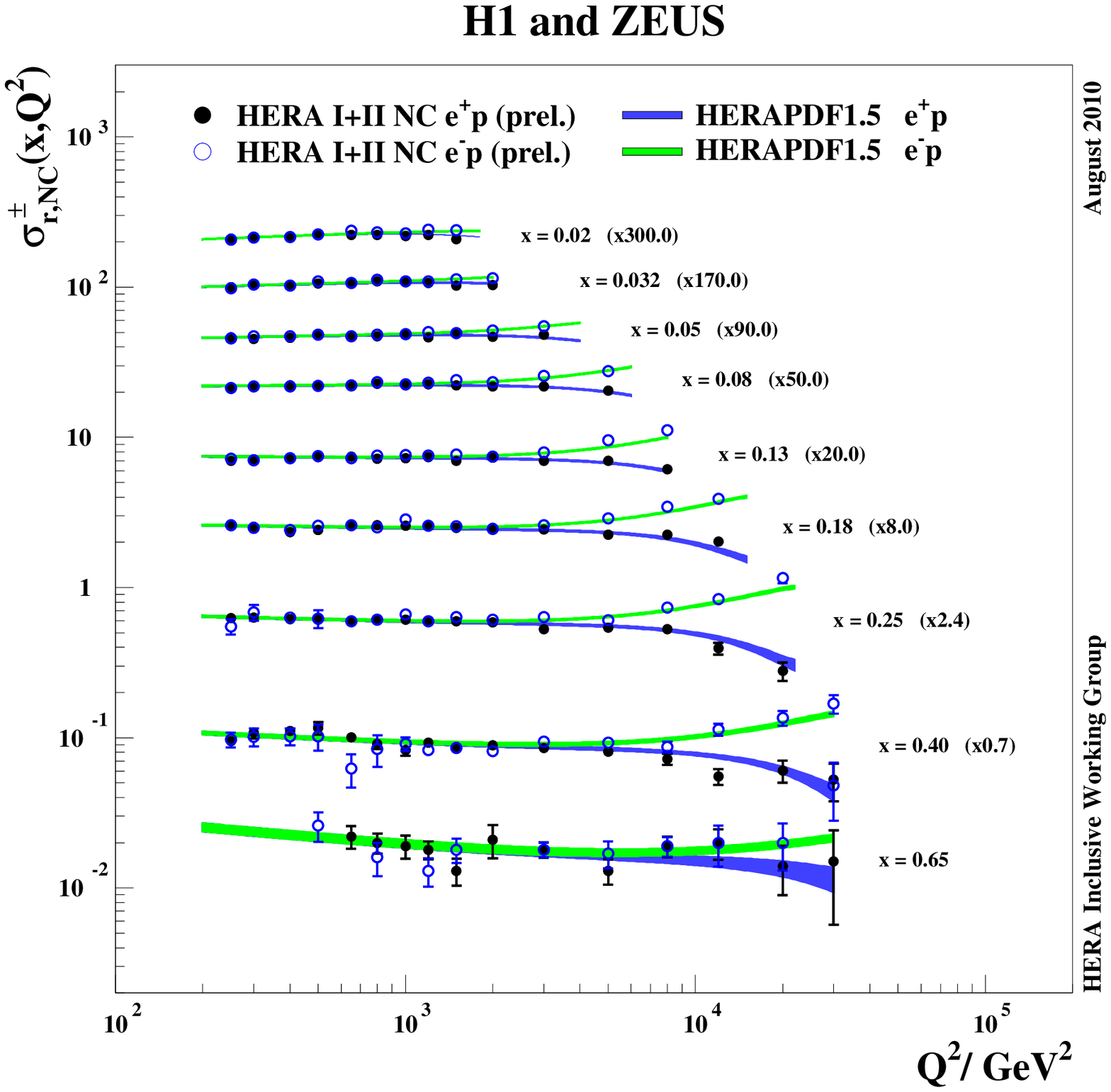,width=0.45\textwidth}~~ &
\psfig{figure=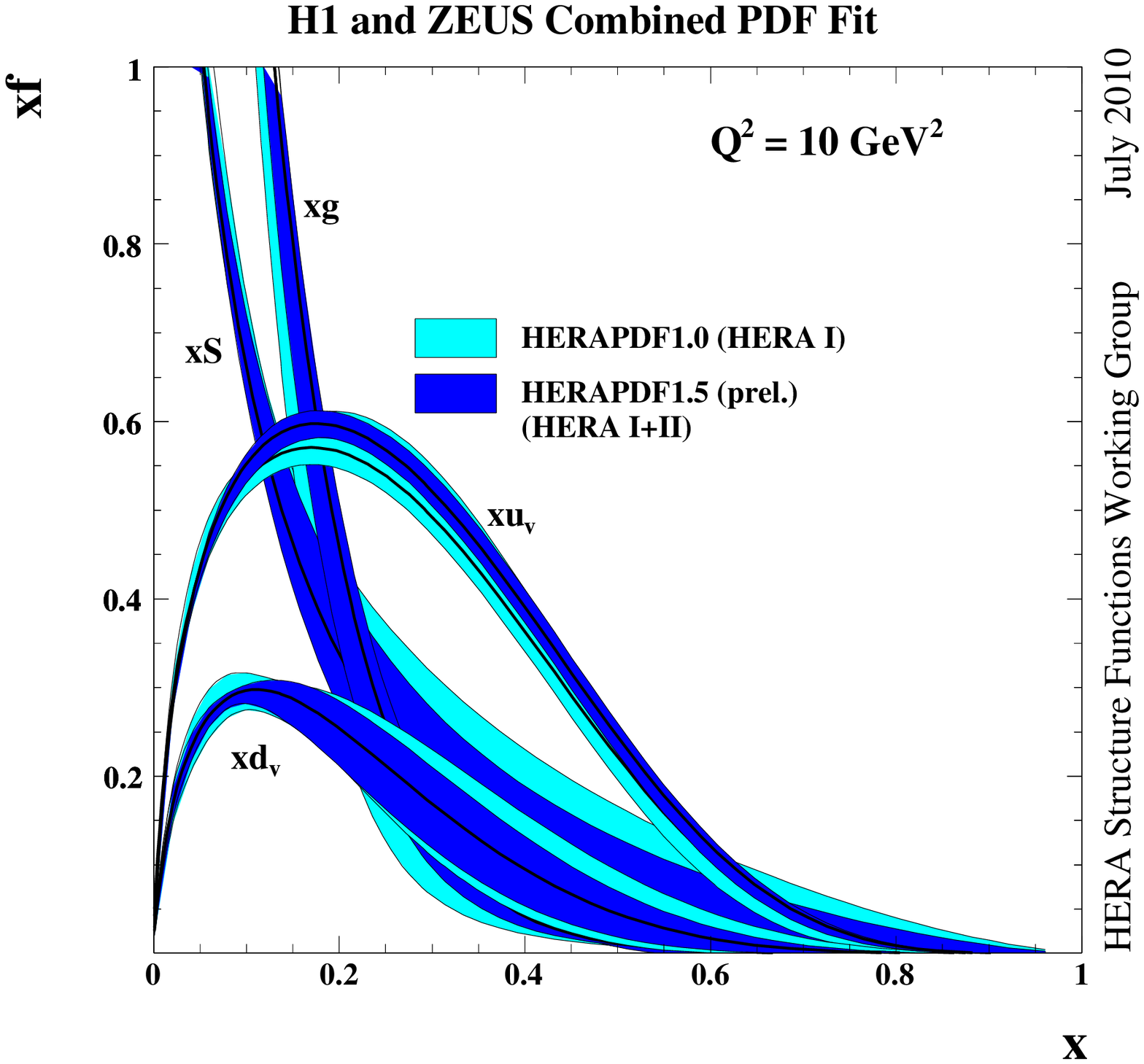,width=0.45\textwidth} \\
\end{tabular} 
\caption{Left: HERA combined data points for the NC $e^{\pm}p$ cross-sections 
as a function of $Q^2$ in bins of $x$, for data from the HERA-I and II run 
periods. The HERAPDF1.5 fit to these data is also shown on the plot.
Right: Parton distribution functions from HERAPDF1.0 and HERAPDF1.5; $xu_v$, 
$xd_v$,$xS=2x(\bar{U}+\bar{D})$ and $xg$ at $Q^2=10~$GeV$^2$.}
\label{fig:herapdf15}
\end{center}
\end{figure}

\begin{figure}[tbp]
\vspace{-2.0cm} 
\begin{center}
\begin{tabular}{ll}
\psfig{figure=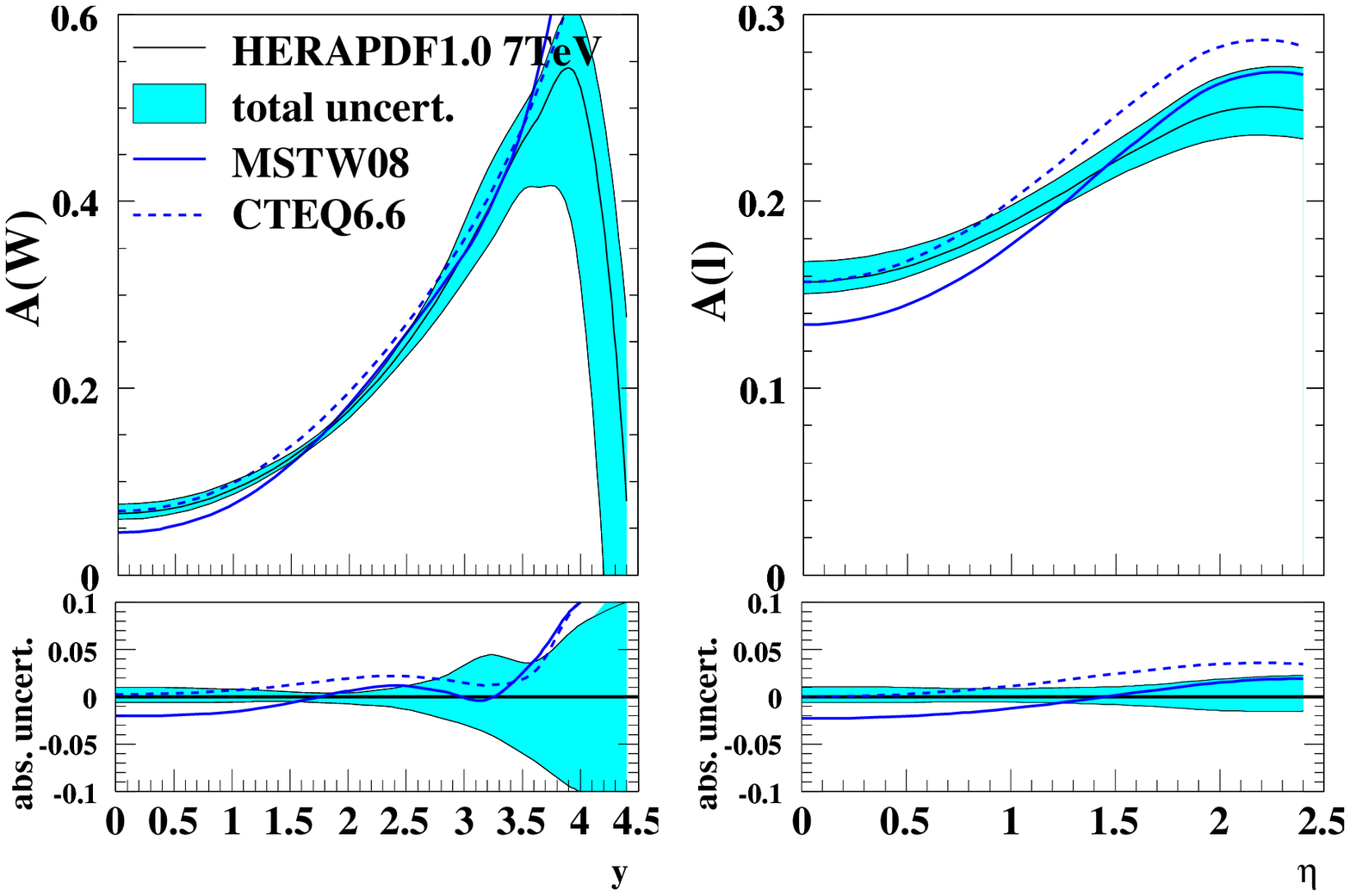,width=0.45\textwidth}~~ &
\psfig{figure=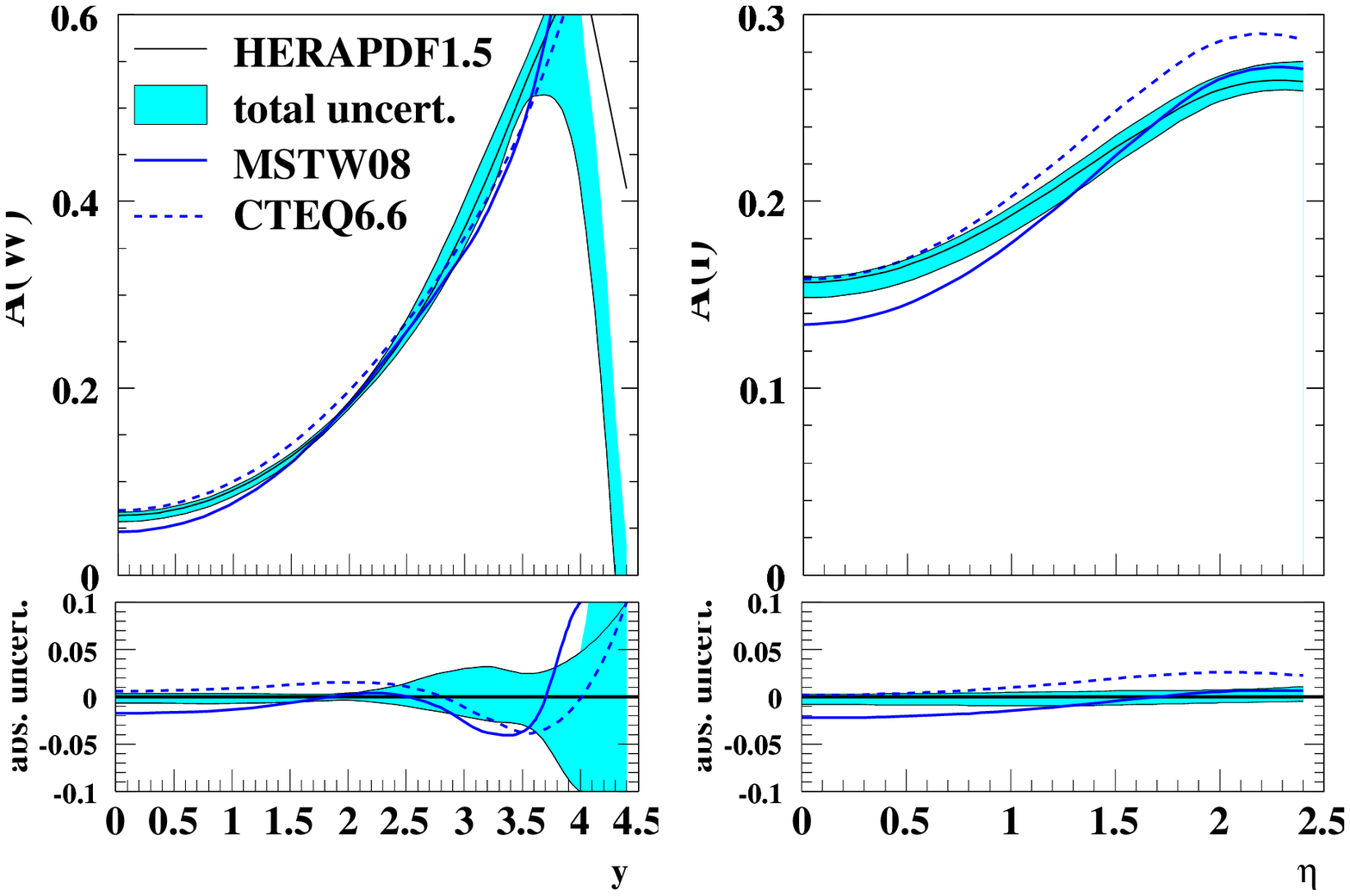,width=0.45\textwidth} \\
\end{tabular} 
\caption{Predictions for the $W$ and lepton asymmetries from HERAPDF1.0 (left) and HERAPDF1.5 (right). The MSTW2008 and CTEQ6.6 PDf predictions are also shown.}
\label{fig:lhc}
\end{center}
\end{figure}

Preliminary H1 data on $NC$ and $CC$ $e^+p$ and $e^-p$ inclusive cross-sections
and published ZEUS data on $NC$ and $CC$ $e^-p$ and $CC$ $e^+p$ data, from 
HERA-II running, have been combined with the HERA-I data to yield an inclsuive 
data set wih improved accuracy at high $Q^2$ and high $x$~\cite{highq2}. 
The HERA-I data set and the new HERA I+II data sets are compared for 
$CC$ $e^-p$ data in Fig.~\ref{fig:ccem1015}. This new data set is used as the 
sole input to  a PDF fit called the HERAPDF1.5 which uses the same formalism 
and assumptions as the HERAPDF1.0 fit~\cite{hiq2study}. 
These fits are superimposed on the corresponding data sets in the figure.
Fig.~\ref{fig:herapdf15} (left) shows the combined data for $NC$ $e^{\pm}p$ 
cross-sections with the HERAPDF1.5 fit superimposed. The parton distribution
functions from HERAPDF1.0 and HERAPDF1.5 are compared in 
Fig.~\ref{fig:herapdf15} (right). The improvement in precision at high $x$ 
is clearly visible.
The impact of these improved data on predictions for the LHC may be gauged from
Fig.~\ref{fig:lhc} which compares predictions for the $W$ and lepton 
asymmetries from the HERAPDF1.0 and HERAPDF1.5 fits. 

\section{Summary}
The status of the combinations of HERA data and the PDF fits based on these 
data have been reviewed. These data form the basis for accurate predictions
of LHC cross-sections.


\begin{footnotesize}

\end{footnotesize}


\end{document}